 \input colordvi
\documentclass[11pt]{article}
\usepackage{amsfonts,amssymb,a4}

\newcommand{\qed}{\hfill\rule{3mm}{3mm}}
\newtheorem{teorema}{Theorem}
\newtheorem{lema}{Lemma}
\newtheorem{corolario}{Corollary}

\makeatletter \@addtoreset{equation}{section} \makeatother
\begin{document}


\voffset=-1.5truecm\hsize=16.5truecm    \vsize=24.truecm
\baselineskip=14pt plus0.1pt minus0.1pt \parindent=12pt
\lineskip=4pt\lineskiplimit=0.1pt      \parskip=0.1pt plus1pt

\def\ds{\displaystyle}\def\st{\scriptstyle}\def\sst{\scriptscriptstyle}


\let\a=\alpha \let\b=\beta \let\chi=\chi \let\d=\delta \let\e=\varepsilon
\let\f=\varphi \let\g=\gamma \let\h=\eta    \let\k=\kappa \let\l=\lambda
\let\m=\mu \let\n=\nu \let\o=\omega    \let\p=\pi \let\ph=\varphi
\let\r=\rho \let\s=\sigma \let\t=\tau \let\th=\vartheta
\let\y=\upsilon \let\x=\xi \let\z=\zeta
\let\D=\Delta \let\F=\Phi \let\G=\Gamma \let\L=\Lambda \let\Th=\Theta
\let\O=\Omega \let\P=\Pi \let\Ps=\Psi \let\Si=\Sigma \let\X=\Xi
\let\Y=\Upsilon


\global\newcount\numsec\global\newcount\numfor
\gdef\profonditastruttura{\dp\strutbox}
\def\senondefinito#1{\expandafter\ifx\csname#1\endcsname\relax}
\def\SIA #1,#2,#3 {\senondefinito{#1#2}
\expandafter\xdef\csname #1#2\endcsname{#3} \else \write16{???? il
simbolo #2 e' gia' stato definito !!!!} \fi}
\def\etichetta(#1){(\veroparagrafo.\veraformula)
\SIA e,#1,(\veroparagrafo.\veraformula)
 \global\advance\numfor by 1
 \write16{ EQ \equ(#1) ha simbolo #1 }}
\def\etichettaa(#1){(A\veroparagrafo.\veraformula)
 \SIA e,#1,(A\veroparagrafo.\veraformula)
 \global\advance\numfor by 1\write16{ EQ \equ(#1) ha simbolo #1 }}
\def\BOZZA{\def\alato(##1){
 {\vtop to \profonditastruttura{\baselineskip
 \profonditastruttura\vss
 \rlap{\kern-\hsize\kern-1.2truecm{$\scriptstyle##1$}}}}}}
\def\alato(#1){}
\def\veroparagrafo{\number\numsec}\def\veraformula{\number\numfor}
\def\Eq(#1){\eqno{\etichetta(#1)\alato(#1)}}
\def\eq(#1){\etichetta(#1)\alato(#1)}
\def\Eqa(#1){\eqno{\etichettaa(#1)\alato(#1)}}
\def\eqa(#1){\etichettaa(#1)\alato(#1)}
\def\equ(#1){\senondefinito{e#1}$\clubsuit$#1\else\csname e#1\endcsname\fi}
\let\EQ=\Eq


\def\\{\noindent}
\let\io=\infty

\def\VU{{\mathbb{V}}}
\def\ED{{\mathbb{E}}}
\def\GI{{\mathbb{G}}}
\def\Tt{{\mathbb{T}}}
\def\C{\mathbb{C}}
\def\LL{{\cal L}}
\def\RR{{\cal R}}
\def\SS{{\cal S}}
\def\NN{{\cal M}}
\def\MM{{\cal M}}
\def\HH{{\cal H}}
\def\GG{{\cal G}}
\def\PP{{\cal P}}
\def\AA{{\cal A}}
\def\BB{{\cal B}}
\def\FF{{\cal F}}
\def\TT{{\cal T}}
\def\v{\vskip.1cm}
\def\vv{\vskip.2cm}
\def\gt{{\tilde\g}}
\def\E{{\mathcal E} }
\def\I{{\rm I}}
\def\0{\emptyset}

\def\tende#1{\vtop{\ialign{##\crcr\rightarrowfill\crcr
              \noalign{\kern-1pt\nointerlineskip}
              \hskip3.pt${\scriptstyle #1}$\hskip3.pt\crcr}}}
\def\otto{{\kern-1.truept\leftarrow\kern-5.truept\to\kern-1.truept}}
\def\arm{{}}
\font\bigfnt=cmbx10 scaled\magstep1

\newcommand{\card}[1]{\left|#1\right|}
\newcommand{\und}[1]{\underline{#1}}
\def\1{\rlap{\mbox{\small\rm 1}}\kern.15em 1}
\def\ind#1{\1_{\{#1\}}}
\def\bydef{:=}
\def\defby{=:}
\def\buildd#1#2{\mathrel{\mathop{\kern 0pt#1}\limits_{#2}}}
\def\card#1{\left|#1\right|}
\def\proof{\noindent{\bf Proof. }}
\def\qed{ \square}
\def\reff#1{(\ref{#1})}
\def\eee{{\rm e}}

\begin{center}
{\LARGE Regions without complex zeros for chromatic polynomials on
graphs with bounded degree} \vskip1.0cm {\large Roberto
Fernandez$^{1}$ and Aldo Procacci$^{1,2}$}

\vskip.5cm

{\footnotesize $^{1}$Labo. de Maths Raphael SALEM, UMR 6085
CNRS-Univ. de Rouen, Avenue de l'Universit\'e, BP.12,
76801 Saint Etienne du Rouvray,  France\\
\vskip.1cm
 $^{2}$
Dep. Matem\'atica-ICEx, UFMG, CP 702, Belo Horizonte MG
30.161-970, Brazil

 email: $^1${\tt Roberto.Fernandez@univ-rouen.fr}; ~
$^2${\tt aldo@mat.ufmg.br}}

\end{center}

\def\be{\begin{equation}}
\def\ee{\end{equation}}
\vskip.5cm

\begin{abstract}
We prove that the chromatic polynomial $P_\mathbb{G}(q)$ of a
finite graph $\mathbb{G}$ of maximal degree $\D$ is free of zeros
for $\card q\ge C^*(\D)$ with
$$ C^*(\D)\;=\; \min_{0<x<2^{1\over \D}-1}\,
{(1+x)^{\D-1}\over x\, [2-(1+x)^\D]}
$$
This improves results by Sokal (2001) and Borgs (2005).
Furthermore, we present a strengthening of this condition for
graphs with no triangle-free vertices.
\end{abstract}

\section{Introduction}

Let $\mathbb{G}=(\mathbb{V},\mathbb{E})$ be a finite graph with
vertex set $\mathbb{V}$, edge set $\mathbb{E}$, and maximum degree
$\D$. For any integer $q$, let $P_\mathbb{G}(q)$ be equal to the
number of proper colorings with $q$ colors of the graph
$\mathbb{G}$, that is colorings such that no two adjacent vertices
of the graph have equal colors. The function $P_\mathbb{G}(q)$ is
a polynomial known as the \emph{chromatic polynomial}, and it
coincides with the partition function of the anti-ferromagnetic
Potts model with $q$ states on $\mathbb{G}$ at zero temperature.
Sokal~\cite{sok01} exploited a well known representation of the
latter which leads to the identity
\begin{equation}
P_\mathbb{G}(q)\;=\;q^{|\mathbb{V}|}\,\Xi_\mathbb{G}(q)\;.
\label{eq:1}
\end{equation}
Here $\Xi_\mathbb{G}(q)$ is the grand canonical partition function
of a ``gas'' whose ``particles'' are subsets $\g\subset
\mathbb{V}$, with cardinality $|\g|\ge 2$,  subjected to a
non-intersection constraint (hard-core interaction) and endowed
with activities $z_\g(q)$ that depend on the topological structure
of $\mathbb{G}$ [see \reff{2.14} below].  Such a hard-core gas
corresponds to an abstract \emph{polymer model}~\cite{grukun71}
whose analyticity properties are the object of the \emph{cluster
expansion} technology~\cite{cam82,bry84,kotpre86,dob96}. The
absolute convergence of the cluster expansion yields the
analyticity of $\log\Xi_\mathbb{G}(q)$ as a function of the
activities and, thus, the absence of zeros of $P_\mathbb{G}$ for
the corresponding \emph{complex} disk in $q$.

At this point, one can make use of any of the available
convergence conditions for the cluster expansion.  Sokal used the
Kotecky-Preiss condition~\cite{kotpre86} which requires the
existence of some $a>0$ such that
\begin{equation}
\sum_{n\ge 2}\eee^{an}\,C_n^q\;\le\; a \label{KP}
\end{equation}
where
\begin{equation}
C_n^q\;=\;\sup_{x\in \mathbb{V}}\sum_{\g\subset
\mathbb{V}:\,\,x\in \g\atop \card\g=n} \bigl|z_\g(q)\bigr|\;.
\label{sup}
\end{equation}
He then combined this  condition with the bound~~\cite{sok01}
\begin{eqnarray}
\label{eq:ar.0}
C_n^q &\le& \left({1\over q}\right)^{n-1} \sup_{v_0\in \mathbb{V}_\mathbb{G}} t_n(\mathbb{G},v_0) \\
&\le& \left({1\over q}\right)^{n-1} t_{n}(\D)
\label{eq:ar.1}
\end{eqnarray}
where $t_n(\mathbb{G},v_0)$ is number of  subtrees of
$\mathbb{G}$, with  $n$ vertices, one of which is $v_0$ and
$t_{n}(\D)$ is the number of $n$-vertex  subtrees in the
$\D$-regular infinite tree containing a fixed vertex. Using
\reff{eq:ar.1}, Sokal proved that $P_\mathbb{G}(q)$ is free of
zeros in the region
\begin{equation}
\card q\;\ge\; C(\D) \label{sok.zero}
\end{equation}
where $C(\D)$ is defined by
\begin{equation}\label{CD}
C(\D)= \min_{a\ge 0} \,\inf\biggl\{\k: \sum_{n=1}^{\infty}
t_n(\D)\Bigl[\frac{\eee^a}{\k}\Bigr]^{n-1}\le \, 1+
a\,\eee^{-a}\biggr\}\;.
\end{equation}
By numerical methods, Sokal obtained rigorous upper bounds on
$C(\D)$ for $2\le \D\le 20$ (see Table 2 in~\cite{sok01} for
$\D\le20$).  He also showed that for large $\D$ there is a finite
limit $\lim_{\D\to \infty}C(\D)/\D=K$ with
\begin{eqnarray}
K&=&\min_{a\ge 0} \,\inf\biggl\{\k: \sum_{n=1}^{\infty}
\frac{n^{n-1}}{n!} \Bigl[\frac{\eee^a}{\k}\Bigr]^{n-1}\le 1+
a\,\eee^{-a}\biggr\}
\label{Ka}\\[8pt]
&=& \min_{a\ge 0}\,
\frac{\exp\Bigl\{a+\ln(1+a\,\eee^{-a})\Bigr\}}{\ln(1+a\,\eee^{-a})}
\label{Kaex}\\[8pt]
&=& 7.963906\ldots\label{eq:2}
\end{eqnarray}
The expression \reff{Ka} is the one given originally
in~\cite{sok01}, where the estimation \reff{eq:2} ---and  the
rigorous bound $K\le 7.963907$--- were obtained through a
computer-assisted calculation. Its identification with \reff{Kaex}
is due to Borgs~\cite{bor05}. Furthermore, this constant $K$ is
such that $C(\D)\le K\D$ for all $\D$, thus yielding, for the
region free of zeroes, the weaker but simpler bound
\begin{equation}
\card q\;\ge\; K \D\;, \label{zero}
\end{equation}
which approaches \reff{sok.zero} in the large-$\D$ regime. The
bound \reff{zero}--\reff{Ka}  can be obtained in a more direct way
simply by combining \reff{KP} with the previously obtained
inequality~\cite{proscoger03}
\begin{equation}
C_n^q\;\le\; {n^{n-1}\over n!}\left[{\D\over q}\right]^{n-1}\;.
\label{acti}
\end{equation}

In this paper we improve these criteria in two different
directions. On the one hand our bounds improve Sokal's results for
graphs for which the maximum degree is the only available
information.  On the other hand, we are able to exploit relations
between vertices with a common neighbor to produce even better
bounds if the graph has no triangle-free vertex (a vertex is
triangle-free if there is no edge linking two of its neighbors).
These improvements have a double source.  First, we strengthen the
convergence criterium \reff{KP} replacing $a$ by $\eee^a-1$ in the
right-hand side (Lemma \ref{lem:1}). Second, we improve the bound
\reff{eq:ar.1} by considering a restricted family of trees (Lemma
\ref{lem:3b}). Both improvements are in fact, related, and amount
to a more careful consideration of an identity due to Penrose
\cite{pen67}.  Our ideas stem from the work reported in \cite{FP},
even when below we produce independent, self-contained proofs.

\section{Results}

Let us introduce some additional notation. Given $v_0\in
\mathbb{V}$, let $d_{v_0}$ be its degree and $\G(v_0)=\{v\in
\mathbb{V}: \{v,v_0\}\in \mathbb{E}\}$ its neighborhood.
Let, for $k=1,\dots \D$,  
\begin{equation}
t_k^{\mathbb{G}}\;=\; \sup_{v_0\in \mathbb{V}\atop d_{v_0}\ge
k}\card{\Big\{U\subset \G(v_0): \,\, |U|=k\,\, {\rm and}\,\,
\{v,v'\}\notin \mathbb{E}\;\forall\, v,v'\in U\Big\}}
\label{wpen2}
\end{equation}
(maximal number of families of $k$ vertices that have a common
neighbor but are not neighbors between themselves).  Consider also
\begin{equation}
\widetilde t_k^{\mathbb{G}}= \sup_{v_0\in \mathbb{V}\atop
d_{v_0}\ge k+1}\max_{v\in \G(v_0)}\card{\Big\{U\subset
\G(v_0)\backslash\{v\}: \,\, |U|=k\,\, {\rm and}\,\,
\{v,v'\}\notin \mathbb{E}\;\forall\, v,v'\in U\Big\}}\label{wpen}
\end{equation}
(same as above but excluding, in addition, one of the neighbors).
We then denote, for $u>0$,
\begin{equation}\label{ZD}
Z_{\mathbb{G}}(u)=1+\sum_{k=1}^{\D} t^{\mathbb{G}}_k\, u^k
\end{equation}
and
\begin{equation}\label{ZD-1}
\widetilde Z_{\mathbb{G}}(u)=1+\sum_{k=1}^{\D-1}\widetilde
t^{\mathbb{G}}_k\, u^k
\end{equation}
Finally, let  $\bar t_{n}(\D)$ be the number of subtrees of the
$\D$-regular infinite tree which of $n$ vertices, containing a
fixed vertex, say $v_0$, identified as the root, and satisfying
the following constraints:
\begin{itemize}
\item[(i)] The maximum number of subsets
of descendants of $v_0$ with fixed cardinality $k$ (with $1\le
k\le \D$) is $t_k^{\mathbb{G}}$.
\item[(ii)] For any vertex $v\neq v_0$, the maximum number of subsets
of descendants of $v$ with fixed cardinality $k$ (with $1\le k\le
\D-1$)  is $\tilde t_k^{\mathbb{G}}$.
\end{itemize}

\begin{teorema}\label{theo:2}
The chromatic polynomial of a finite graph $\mathbb{G}$ of maximal
degree $\D$ is free of zeros for
\begin{equation}
\card q\;\ge\; C^*_\mathbb{G} \label{eq:3b}
\end{equation}
with
\begin{eqnarray}
C^*_\mathbb{G}&=&\min_{a\ge 0} \,\inf\biggl\{\k:
\sum_{n=1}^{\infty} \bar
t_n(\D)\Bigl[\frac{\eee^a}{\k}\Bigr]^{n-1}\le \,
2-\eee^{-a}\biggr\}
\label{Ka.1b}\\[8pt]
&=& \min_{a>0}\, \eee^a\,{\tilde Z_\mathbb{G}\Big(Z_\mathbb{G}^{-1}( 2-\eee^{-a})\Big)\over Z_\mathbb{G}^{-1}( 2-\eee^{-a})}\label{Ka.2}\\
&=&
\min_{0 <x<Z_\mathbb{G}^{-1}(2)}\, {\widetilde
Z_{\mathbb{G}}\big(x\big)\over [2-Z_\mathbb{G}(x)] x}
\label{eq:4b}
\end{eqnarray}
\end{teorema}

The numbers $t_k^{\mathbb{G}}$ and $\widetilde t_k^{\mathbb{G}}$
depend on the graph structure. They depend on the presence
of neighbors of a point that are themselves neighbors, that is on
the existence of triangle diagrams in the graph. In fact, they satisfy
the inequalities
\begin{equation}
\Delta\,\delta_{k\,1}\;\le\;t_k^{\mathbb{G}}\;\le\; {\D \choose k} \quad,\quad
(\Delta-1)\,\delta_{k\,1}\;\le\;\widetilde t_k^{\mathbb{G}}\;\le\;{\D-1 \choose k}\;.
\label{inv.1}
\end{equation}
The lower bound ($\delta_{k\,1}=$ if $k=1$ and zero otherwise) corresponds
to the complete graph with $\Delta+1$ vertices.  This is the graph with the largest
possible number of triangle diagrams per vertex, and hence for which
the improvement contained in the previous theorem is maximal.  In this case,
$Z_{\mathbb{G}_{\rm cpl}}(u)= 1+\D u$,
$\widetilde Z_{\mathbb{G}_{\rm cpl}}(u)= 1+(\D-1)u$ and
a straightforward calculation
shows that
\begin{equation}
C^*_\mathbb{G_{\rm cpl}} \;=\; \frac{(\D-1)^2}
{3\D-1-2\sqrt{2\D^2-\D}}\;.
\label{inv.3}
\end{equation}
We check that $C^*_\mathbb{G_{\rm cpl}} (\D)/\D$ is an increasing function of $\D$ and
\begin{equation}
\frac{C^*_\mathbb{G_{\rm cpl}}}{\D} \;\mathop{\nearrow}\limits_{\D\to\infty}\;
\frac{1}{3-2\sqrt 2}\,\D\;\approx\; 5.83\,\D\;.
\label{inv.6}
\end{equation}

The upper bounds in \reff{inv.1} corresponds to graphs with
a triangle-free vertex of degree $\D$.  It is simple to see,
for instance from \reff{Ka.1b}, that the use of these upper
bounds yields a worst-scenario estimation for any graph.
In this case
$Z_{\mathbb{G}}(u)= (1+u)^\D$,
$\widetilde Z_{\mathbb{G}}(u)= (1+u)^{\D-1}$ and $\bar t_{n}(\D)=
t_n(\D)$.  In this way, Theorem
\ref{theo:2} yields the following corollary for general graphs with maximum
degree $\Delta$.

\begin{corolario}\label{theo:1}
The chromatic polynomial of a finite graph
$\mathbb{G}=(\mathbb{V},\mathbb{E})$ of maximal degree $\D$ is
free of zeros for
\begin{equation}
\card q\;\ge\; C^*(\D) \label{eq:3}
\end{equation}
with
\begin{eqnarray}
C^*(\D)&=&\min_{a\ge 0} \,\inf\biggl\{\k: \sum_{n=1}^{\infty}
t_n(\D)\Bigl[\frac{\eee^a}{\k}\Bigr]^{n-1}\le \, 2-e^{-a}\biggr\}
\label{Ka.1}\\[8pt]
\label{eq:4.-1} &=& \min_{a>0}\,
{\eee^a (2-\eee^{-a})^{1-{1\over \D}}\over (2-\eee^{-a})^{1\over \D}-1}\label{Ka.2b}\\
&=&
\min_{0<x<2^{1\over \D}-1}\, {(1+x)^{\D-1}\over x\,[2-(1+x)^\D]}
\label{eq:4}
\end{eqnarray}
\end{corolario}
The equality of \reff{Ka.1} and \reff{eq:4.-1}/\reff{eq:4} is a
generalization of Borgs' identity \cite{bor05} connecting
\reff{Ka} with \reff{Kaex}. In fact, the identity between
\reff{Ka.1} and \reff{eq:4} can be equally well applied to
\reff{CD} just replacing the factor $2-\eee^{-a}$ with the factor
$1+a\eee^{-a}$. This yields the following alternative expression
for Sokal's constant \reff{CD}:
\begin{equation}
C(\D)= \min_{a>0} {\eee^a (1+a\eee^{-a})^{1-{1\over \D}}\over
(1+a\eee^{-a})^{1\over \D}-1}\;.
\label{inv.2}
\end{equation}

The function $C^*(\D)/\D$ increases with
$\D$; thus, \reff{eq:3} implies the following rougher
bound.
\begin{corolario}\label{coro:1}
The chromatic polynomial of a finite graph
$\mathbb{G}=(\mathbb{V},\mathbb{E})$ of maximal degree $\D$ is
free of zeros for
\begin{equation}
\card q\;\ge\; K^*\,\D \label{eq:5}
\end{equation}
with
\begin{equation}
K^*\;=\; \lim_{\D\to\infty}{ C^*(\D)\over \D} \;=\; \min_{a>0}\,
\frac{\exp\bigl\{a +\ln[2-\eee^{-a}]\bigr\}} {\ln[2-\eee^{-a}]}=
\min_{1<y<2}\, \frac{y} {(2-y)\ln y}\;. \label{eq:6}
\end{equation}
\end{corolario}
The bound \reff{eq:5}--\reff{eq:6} is a strengthening of
\reff{zero}--\reff{Ka}/\reff{Kaex}. For example for $y=1.3702$
(that is, $a\approx 0,46235$), we get $ K^*\;\le\;
6.907\ldots $.

Table \ref{tab:1} presents some examples of the different
estimations discussed in this paper.

\begin{table}[h]
\begin{center}
\begin{tabular}{|c||r|r||r|r|}
\hline
 &\multicolumn{2}{c||}{General graph}&
\multicolumn{2}{c|}{Complete graph}\\
\cline{2-5}
${}^{\textstyle\D}$ & \cite{sok01} & \reff{eq:3} & \reff{eq:3b}/\reff{inv.3} & Exact\\
\hline
\hline
2 & 13.23 & 10.72 & 9.90 & 2\\
3 & 21.14 & 17.57 & 15.75 & 3\\
4 & 29.08 & 24.44 & 21.58 & 4\\
6 & 44.98 &  38.24 & 33.24 & 6 \\
Any & 7.97$\D$ & 6.91$\D$ & 5.83$\D$ & $\D$\\
\hline
\end{tabular}
\end{center}
\caption{\label{tab:1}Comparison of the different criteria for graphs of maximum degree
$\D$.  Each entry gives the value $C_{\mathbb{G}}$ such that the chromatic
polynomial is free of zeros for $\left|q\right|> C_{\mathbb{G}}$}
\end{table}


%

\section{Proofs}

\subsection{The basic lemmas}

Theorem \ref{theo:1} is an immediate consequence of
the following three lemmas.

\begin{lema}\label{lem:1}
Consider the lattice gas with activities $\{z_\g(q):\g\subset
\mathbb{V}\}$ described above. Then its cluster expansion
converges if $q>\eee\D$ and there exists $a>0$ such that
\begin{equation}
\sum_{n\ge 2}\eee^{an}\,C_n^q\;\le\; \eee^a - 1\;. \label{FP}
\end{equation}
\end{lema}

\begin{lema}\label{lem:3b}

Consider the lattice gas with activities $\{z_\g(q):\g\subset V\}$
described above. The activities satisfy the bounds
\begin{equation}
C_n^q\;\le\;   \left({1\over q}\right)^{n-1}\bar t_n(\D)
\label{newr}
\end{equation}
\end{lema}

 \begin{lema}\label{lem:3}
The formal power series
\begin{equation}\label{sTbarb}
\overline T(x)\;=\;\sum_{n=1}^\infty  \bar t_{n}(\D)\, x^n
\end{equation}
converges for all  $x\in [0,R)$ where
$$
R\;=\;\sup_{u\ge 0}{u\over \widetilde Z_{G}(u)}
$$
and
\begin{equation}
\sup\Bigl\{ 0<x< R: \frac{1}{x}\,\overline T(x)\le b \Bigr\} \;=\;
{Z_G^{-1}( b)\over \widetilde Z_{G}\big(Z_G^{-1}( b)\big)}.
\label{eq:34b}
\end{equation}
\end{lema}

Condition \reff{FP} follows from general results on cluster
expansions fully developed in \cite{FP}.  For completeness, we
present a simple direct proof in the sequel, which, as the work
in~\cite{FP}, crucially depends on an identity due to Penrose
\cite{pen67}. The bound (\ref{newr}) is an  improvement respect to the bound (\ref{eq:ar.1}) only for graphs with no
triangle free vertices.
Finally, Lemma \ref{lem:3} is also a
simplified version of the argument in \cite{FP}.

Before turning to the proof of these lemmas we discuss some needed notions
of the theory of cluster expansions.

\subsection{Activities and polymer expansion}

We start by summarizing the fundamental expressions.  The reader
can consult \cite{sok01} for its derivation.  The activities
$z_\g$ of the hard-core partition function $\Xi_{\mathbb{G}}(q)$
depend on the graphs $\mathbb{G}_\g=(\g,\mathbb{E}_\g)$ obtained
restricting of the original graph $\mathbb{G}$ to the vertex set
$\g$ (that is, $\mathbb{E}_\g=\bigl\{\{x,y\}\in \mathbb{E}: x\in
\g \,\,{\rm and}\,\,y\in\g\bigr\}$).  In the sequel, given graphs
$G'=(V_{G'},E_{G'})$ and $G=(V_G,E_G)$, we say that $G'$ is a
subgraph of $G$, and we write $G'\subset G$, if $V_{G'}\subset
V_G$ and  $E_{G'}\subset E_G$.

Let us denote
\begin{equation}
\MM_\mathbb{G}=\Bigl\{\g\subset \mathbb{V}: \,|\g|\ge 2\,,\,\,
\mathbb{G}_\g \mbox{ connected}\Bigr\} \label{eq:7}
\end{equation}
(the set of \emph{monomers}).  Then,
\begin{equation}
\Xi_{\mathbb{G}}(q)\;=\;\sum_{n\ge 1}\, \sum_{\{\g_1,\dots
,\g_n\}:\,\g_i\in\MM_\mathbb{G}\atop \g_i\cap\g_j=\0}
z_{\g_1}(q)\cdots z_{\g_n}(q) \label{2.6}
\end{equation}
with
\begin{equation}
z_\g\;=\; {1\over q^{\card\g-1}}\sum\limits_{G'\subset
\mathbb{G}_\g:\, V_{G'}=\g\atop G'\; {\rm connected}}
 (-1)^{\card{E_{G'}}}
\label{2.14}
\end{equation}
Note that the sum above run over all spanning subgraphs of
$\mathbb{G}_\g$. The logarithm of this partition function leads to
the cluster or polymer expansion for this model. For each ordered
family $(\g_1,\ldots,\g_n)$ of monomers let $g(\g_1,\ldots,\g_n)$
be the graph with vertex set
$V_{g(\g_1,\ldots,\g_n)}=\{1,2,\dots,n\}$ and edge set
$E_{g(\g_1,\ldots,\g_n)}=\bigl\{\{i,j \}\subset
V_{g(\g_1,\ldots,\g_n)}: \, \g_i\cap \g_j\neq \emptyset\bigr\}$.
In the sequel we will denote shortly $\I_n\doteq\{1,2,\dots,n\}$.

The hard-core lattice gas cluster expansion, or {\it Mayer series}
(see e.g. \cite{rue69} or \cite{cam82} and references therein), is
the \emph{formal} series
\begin{equation}
\Sigma_{\mathbb{G}}(q)\;=\; \sum_{n= 1}^\infty{1\over n!}
\sum_{(\g_{1},\dots ,\g_{n})\in(\MM_\mathbb{G})^n}
\phi^T\bigl[g(\g_1,\dots,\g_n)\bigr]\,z_{\g_1}(q)\dots z_{\g_n}(q)
\label{6}
\end{equation}
with
\begin{equation}
\phi^T\bigl[g(\g_1,\dots,\g_n)\bigr]\;=\;
\cases{\sum\limits_{G'\subset g(\g_1,\dots,\g_n)\atop  G'\ {\rm
connected},\; V_{G'}=\I_n}
 (-1)^{\card{E_{G'}}} &if $n\ge 2$\,\, and $g(\g_1,\ldots,\g_n)\, \mbox{ connected}$\cr\cr
1 &if $n=1$\cr\cr 0 &otherwise } \label{7}
\end{equation}
We shall prove that under condition \reff{FP}, this formal series
converges absolutely. Then
$\Sigma_\mathbb{G}(q)=\ln\Xi_{\mathbb{G}}(q)$ is finite and  the
chromatic polynomial has no zeros.

\subsection{Labeled trees and the Penrose identity}
Expressions \reff{2.14} and \reff{7} ask for the study of
\begin{equation}
S_{\GG} \;=\; \sum\limits_{G'\subset \GG,\, V_{G'}=\I_n \atop G'\;
{\rm connected}}
 (-1)^{\card{E_{G'}}}
\label{eq:9}
\end{equation}
for a connected graph $\GG$ with vertex set $V_\GG=\I_n$ and edge
set $E_\GG$. Penrose~\cite{pen67} produced a crucial identity
relating $S_{\GG}$ to the cardinality of a certain subset of the
set of all spanning trees of $\GG$.

Let  $\TT_\GG$ be the family of all possible trees with vertex set
$\I_n$ which are subgraphs of $\GG$. In other words  $\TT_\GG$ is
the set of spanning trees of $\GG$. For any $\t\in \TT_\GG$ let us
identify the  vertex $1$ as the \emph{root} of $\t$. So we  regard
the trees of $\TT_\GG$ always as rooted in the vertex $1$.

Let $\tau \in \TT_\GG$ with edge set $E_\t$ and, of course, vertex
set $V_\t=V_\GG=\{1,\dots,n\}$. For each vertex $i\in V_\t$, let {
$d_\t(i)$} be the tree distance of the vertex $i$ to the root $1$,
and let { $i'_\t\in V_\t$} be the unique vertex such that
$\{i'_\t,i\}\in E_\t$ and $d(i'_\t)=d(i)-1$. The vertex  $i'_\t$
is called the predecessor of $i$ and conversely $i$ is called the
descendant of $i'_\t$. The number $d_\t(i)$ is called the {\it
generation number} of the vertex $i$.

Let now $p$ be the map that to each tree $\tau\in \TT_\GG$
associates the graph $p(\tau)\subset \GG$ with vertex set $\I_n$
formed by adding (only once) to $\tau$ all edges $\{i,j\}\in
E_{\GG}\setminus E_\tau$ such that either:
\begin{itemize}
\item[(p1)]  $d_\t(i)=d_\t(j)$ (edges between vertices of the same generation), or
\item[(p2)]  $d_\t(j)=d_\t(i)-1$ and $j>i_\t'$ (edges between vertices with generations differing by one).
\end{itemize}

Then the set $\PP_\GG\subset \TT_\GG$  of \emph{Penrose trees} is
defined as
\begin{equation}\label{Pentree}
P_{\GG}\;=\; \bigl\{ \tau\in \TT_\GG :\,\,p(\tau)=\tau \bigr\}\;.
\end{equation}
Thus, a tree $\t\in \TT_\GG$ is a Penrose tree, i.e. $\t\in
\PP_\GG$, if and only if the following two conditions are both
satisfied:
\begin{itemize}

\item[(t1)]  if two vertices $i$ and $j$ of $\t$ have the same generation number (i.e.
$d_\t(i)=d_\t(j)$), then $\{i,j\}\not\in E_\GG$;
\item[(t2)]  If two vertices $i$ and $j$ of $\t$ are such that $d_\t(j)=d_\t(i)-1$ and $j>i_\t'$, then $\{i,j\}\not\in E_\GG$;
\end{itemize}
\\

Penrose identity simply says that
\begin{equation}
S_\GG\;=\; (-1)^{\card{V_\GG}-1}\,\bigl|\PP_\GG\bigr|\;.
\label{eq:10}
\end{equation}
So, by (\ref{eq:10}), $|S_\GG|$ is just the cardinality of the set
of Penrose trees of $\GG$, and, since $\PP_\GG\subset\TT_\GG$, one
obtains immediately the well known bound
\begin{equation}
\card{S_\GG}\;\le \; \card{\TT_\GG}\; \label{eq:10.1b}
\end{equation}
The inequality \reff{eq:10.1b} is the so called tree-graph bound
which e.g. easily implies the bound \reff{eq:ar.1}.

To obtain our new estimates contained in lemmas 1-4, it is crucial
a new and improved bound  on the factor $|S_\GG|$ (the inequality
(\ref{eq:10.1}) below).  For that, we consider another  family of
spanning trees $\tau$ of $\GG$ which is larger than $\PP_\GG$ but
smaller than $\TT_\GG$. The definition of such intermediate family
is obtained from the definition of $\PP_\GG$ above by ignoring
condition (t2) and keeping only the part of  condition (t1)
referring to descendants of the same predecessor. That is, let us
define the subset $\overline \PP_\GG$ of $\TT_\GG$ formed by all
{\it weakly Penrose trees} of $\GG$ as follows. A tree $\t\in
\TT_\GG$ is a weakly Penrose tree, i.e. $\t\in \overline \PP_\GG$
if and only if the following condition is satisfied:

\begin{itemize}
\item[($\bar{\rm t}$1)]  if two vertices $i$ and $j$ of $\t$ are descendants of the same predecessor (i.e. $i'_\t=j'_\t$), then $\{i,j\}\not\in E_\GG$;
\end{itemize}
Note that $(\rm t1)$ implies $(\bar{\rm t}$1), since any two
vertices $i$ and $j$ in a tree $\t$ which are descendants of the
same predecessor have the same generation number. In conclusion,
with this definitions we have $\PP_\GG\subset \overline
\PP_\GG\subset \TT_\GG$ and thus
\begin{eqnarray}\label{eq:10.1}
\card{S_\GG}\;&\le &\; \card{\overline \PP_\GG}\\
&\le &\; \card{\TT_\GG}\;\nonumber
\end{eqnarray}

\subsection{Proof of Lemma \protect\ref{lem:1}}
Let us denote $\r_\g=\card{z_\g}$.  We shall prove that if
\reff{FP} is satisfied, the series of absolute values
\begin{equation}
\Sigma^*_{\mathbb{G}}(q)\;=\; \sum_{n= 1}^\infty{1\over n!}
\sum_{(\g_{1},\dots ,\g_{n})\in(\MM_\mathbb{G})^n}
\Bigl|\phi^T\bigl[g_{\{\g_1,\dots,\g_n\}}\bigr]\Bigr|\,\r_{\g_1}\dots
\r_{\g_n} \label{eq:14}
\end{equation}
is finite.  Let us denote $B_n$ the $n$-th term of the sum.  We
see that
\begin{equation}
B_1\;=\; \sum_{\g\in\MM_\mathbb{G}} \r_\g \;\le\; \card
{\mathbb{V}} \,\sup_{x\in \mathbb{V}}
\sum_{\g\in\MM_\mathbb{G}\atop \g\ni x} \r_\g \;=\;
 \card {\mathbb{V}} \sum_{s\ge 2} C^q_s\;.
\label{eq:ex}
\end{equation}

Let us bound $B_n$, $n\ge 2$. Given \reff{7} and the Penrose
identity \reff{eq:10},
\begin{equation}
B_n\;=\; {1\over n!}\sum_{\tau\in\TT_n} w(\tau) \label{eq:15}
\end{equation}
with
\begin{equation}
w(\t)\;=\;\sum_{(\g_{1},\dots ,\g_{n})\in (\MM_\GG)^n\atop \t\in
\PP_{g(\g_{1},\dots ,\g_{n})}} \r_{\g_1}\ldots \r_{\g_n}\;.
\label{eq:16}
\end{equation}

Using in the sum the weaker condition $\t\in \bar
\PP_{g(\g_{1},\dots ,\g_{n})}$ we obtain the bound
\begin{equation}
w(\t)\;\le\;\overline w(\t)\;:=\; \sum_{(\g_{1},\dots ,\g_{n})\in
(\MM_\GG)^n\atop \t\in \overline\PP_{g(\g_{1},\dots ,\g_{n})}}
\r_{\g_1}\ldots \r_{\g_n} \label{eq:17}
\end{equation}

It is clear that in the sum in the last right-hand side no two
monomers labelling descendants of the same vertex can intersect.
We now estimate $\overline w(\t)$. For each $\tau\in\TT_n$ let us
denote $d_1,\ldots,d_n$ the coordination numbers (degree,
incidence number) of its vertices. Each $d_i$ is the number of
links having vertex $i$ at one endpoint (thus $d_i-1$ is the
number of descendants of $i$); $1\le d_i\le n-1$ and $\sum_{i=1}^n
d_i=2n-2$. Then we have the following

\begin{lema}\label{lem:4}
For each $\tau\in\TT_n$,
\begin{equation}
\overline w(\t)\;\le\; \sum_{\g_1\in \MM_\mathbb{V}\atop |\g_1|\ge
d_1} {|\g_1|\choose d_1}\,d_1!\,\r_{\g_1}\,
\prod_{i=2}^{n}\biggl[\sup_x\sum_{\g_i\in
\MM_\mathbb{V}\,,\;\g_i\ni x\atop |\g_i|\ge d_i-1} {|\g_i|\choose
d_i-1}(d_i-1)!\, \r_{\g_i}\biggr] \label{14}
\end{equation}
\end{lema}

\proof  The proof follows the strategy introduced in~\cite{cam82}.
The tree is successively ``defoliated" by summing over the labels
of the leaves; this produces some of the factors in the right-hand
side of \reff{14} times the weight of a smaller tree.  While the
idea is simple, its inductive formalization requires some
notation.  Let's partition $\{1,\ldots,n\}=I_0\cup
I_1\cup\cdots\cup I_r$ where $I_i$ is the family of vertices of
the $i$-th generation and $r$ the maximal generation number in
$\tau$.  Recall that the unique vertex of $\t$ of the zero
generation is by definition the root, so we have $I_0=\{1\}$.

We also introduce the ``inflated'' activities
\begin{equation}
\widetilde \r_{\g_i} \;=\; \r_{\g_i}\,{|\g_i|\choose
\ell_i}\,\ell_i!\, \ind{\card{\g_i}\ge\ell_i}\;, \label{eq:18}
\end{equation}
where $\ell_i$ is the number of descendants of the vertex $i$,
namely $\ell_1=d_1$ and $\ell_i=d_i-1$ if $i>1$.  The inductive
argument applies to the following expression which is obtained by
reordering the sum in \reff{eq:17}:
\begin{eqnarray}
\overline w(\tau) &=&
\sum_{\g_{I_0}\in(\MM_\mathbb{G})^{|I_0|}}\r_{\g_{I_0}}
\sum_{\g_{I_1}\in(\MM_\mathbb{G})^{|I_1|}}
C(\g_{I_0},\g_{I_1})\,\r_{\g_{I_2}} \;\cdots\nonumber\\
&&\qquad\cdots\sum_{\g_{I_{r-1}}\in(\MM_\mathbb{G})^{|I_{r-1}|}}
C(\g_{I_{r-2}},\g_{I_{r-1}})\,\r_{\g_{I_{r-1}}}
\sum_{\g_{I_r}\in(\MM_\mathbb{G})^{|I_r|}}
C(\g_{I_{r-1}},\g_{I_r})\,\widetilde\r_{\g_{I_r}}\;. \label{eq:19}
\end{eqnarray}
We are denoting $\g_{I_k}=(\g_j)_{j\in I_k}$ and
$\r_{\g_{I_k}}=\prod_{j\in I_k}\r_{\g_j}$.  At this initial step
of the argument, the tilde in the activities of the last
generation is for free because it involves leaves, i.e.\ vertices
with $\ell_i=0$. The factors $C(\g_{I_{k-1}},\g_{I_k})$ embody
condition $(\overline{\rm t}1)$ which relates only consecutive
generations. To write them in detail we further partition each
$I_k$ according to predecessors.  If we decompose $ I_r=
\cup_{i\in I_{r-1}} I^{(i)}_r$, with $I^{(i)}_r$ being the family
of $\ell_i$ descendants of $i$, we have
\begin{equation}
C(\g_{I_{r-1}},\g_{I_r}) \;=\; \prod_{i\in I_{r-1}}
\biggl[\prod_{1\le j\le\ell_i}
\ind{\g_i\cap\g_{i_j}\neq\emptyset}\biggr] \biggl[\prod_{1\le j<k
\le\ell_i} \ind{\g_{i_j}\cap\g_{i_k}=\emptyset}\biggr]
\label{eq:20}
\end{equation}
where we denote $i_1,\ldots,i_{\ell_i}$ the descendants of $i$.

To trigger the induction, we perform the last sum in \reff{eq:19}:
\begin{equation}
\sum_{\g_{I_r}\in(\MM_\mathbb{G})^{|I_r|}}
C(\g_{I_{r-1}},\g_{I_r})\,\widetilde\r_{\g_{I_r}}\; =\;
\prod_{i\in I_{r-1}}\biggl[
\sum_{(\g_{i_1},\ldots,\g_{i_{\ell_i}})\in(\MM_\mathbb{G})^{\ell_i}\atop
\g_{i_j}\cap\g_i\neq\emptyset\,,\; \g_{i_j}\cap\g_{i_k}=\emptyset}
\widetilde\r_{\g_{i_1}}\cdots
\widetilde\r_{\g_{i_{d_i}}}\biggr]\;. \label{eq:21}
\end{equation}
As the sets $\g_{i_j}$ are disjoint, they must intersect $\g_i$ at
$\ell_i$ \emph{different} points.  These points can be chosen in
$\card{\g_i}(\card{\g_i}-1)\cdots (\card{\g_i}-\ell_i+1)$ ways.
Therefore
\begin{equation}
\sum_{\g_{I_r}\in(\MM_\mathbb{G})^{|I_r|}}
C(\g_{I_{r-1}},\g_{I_r})\,\widetilde\r_{\g_{I_r}}\; \le\;
\prod_{i\in I_{r-1}}\biggl[ {|\g_i|\choose \ell_i}\,\ell_i!\,
\ind{\card{\g_i}\ge\ell_i} \prod_{1\le j\le \ell_i}
\Bigl[\sup_{x\in \mathbb{V}} \sum_{ \g_{i_j}\in\MM_\mathbb{G}\atop
\g_{i_j}\ni x} \widetilde\r_{\g_{i_j}}\Bigr]\biggr] \label{eq:22}
\end{equation}
and
\begin{equation}
\r_{\g_{I_{r-1}}}\,\sum_{\g_{I_r}\in(\MM_\mathbb{G})^{|I_r|}}
C(\g_{I_{r-1}},\g_{I_r})\,\widetilde\r_{\g_{I_r}}\; \le\; \Bigl[
\prod_{i\in I_{r-1}} \widetilde\r_i\Bigr] \prod_{j\in I_{r}}
\Bigl[\sup_{x\in \mathbb{V}} \sum_{ \g_{i_j}\in\MM_\mathbb{G}\atop
\g_{i_j}\ni x} \widetilde\r_{\g_{i_j}}\Bigr]\;. \label{eq:23}
\end{equation}
Applying this inequality to \reff{eq:19}, we obtain
\begin{equation}
\overline w(\tau) \;\le\;
\biggl[\sum_{\g_{I_0}\in(\MM_\mathbb{G})^{|I_0|}}\r_{\g_{I_0}}
\;\cdots \sum_{\g_{I_{r-1}}\in(\MM_\mathbb{G})^{|I_{r-1}|}}
C(\g_{I_{r-2}},\g_{I_{r-1}})\,\widetilde\r_{\g_{I_{r-1}}}\biggr]
\prod_{j\in I_{r}} \Bigl[\sup_{x\in \mathbb{V}} \sum_{
\g_{j}\in\MM_\mathbb{G}\atop \g_{j}\ni x}
\widetilde\r_{\g_{j}}\Bigr]\;. \label{eq:24}
\end{equation}
The first square bracket has exactly the form of the right-hand
side of \reff{eq:19} but involving one less generation.
Inductively we therefore obtain
\begin{equation}
\overline w(\tau) \;\le\;
\biggl[\sum_{\g_{I_0}\in(\MM_\mathbb{G})^{|I_0|}}
\widetilde\r_{\g_{I_0}} \prod_{1\le k\le r} \prod_{j_k\in I_{k}}
\Bigl[\sup_{x\in \mathbb{V}} \sum_{ \g_{j_k}\in\MM_\mathbb{G}\atop
\g_{j_k}\ni x} \widetilde\r_{\g_{j_k}}\Bigr]\;. \label{eq:25}
\end{equation}
This is, precisely, the bound \reff{14}.$\qed$
\bigskip

The bound provided by the preceding lemma is only a function of
the coordination numbers $d_1,\ldots,d_n$ of $\tau$.  Thus, in
\reff{eq:15} we can combine it with Cayley  formula [the number of
trees with such coordination numbers is ${n-2 \choose
d_1-1\,\ldots\, d_n-1}$] to obtain
\begin{equation}
B_n\;\le\; \frac{\card{\mathbb{V}}}{n(n-1)} \sum_{d_1,...,d_n\ge
1\atop\sum d_i=2n-2} d_1 \, F(d_1)\,\prod_{i=2}^n F(d_i-1)
\label{eq:26}
\end{equation}
with
\begin{equation}
F(\ell) \;=\; \sup_x\sum_{\g\in \MM_\mathbb{G}\,,\;\g\ni x\atop
|\g|\ge \ell} {|\g|\choose \ell}\,\ell!\, \r_{\g}\;. \label{eq:27}
\end{equation}
To benefit somehow from the restriction $\sum d_i=2n-2$ we resort
to a trick used in~\cite{prosco99}, which consists in multiplying
and dividing by $\alpha^{n-1}=\alpha^{d_1+(d_2-1)\cdots
+(d_n-1)}$, where $\alpha>0$ is left arbitrary.
\begin{equation}
B_n\;\le\; \frac{\card {\mathbb{V}}}{n(n-1)\,\alpha^{n-1}}
\sum_{d_1\ge 1} d_1 \, F(d_1)\,\alpha^{d_1}\, \prod_{i=2}^n
\Bigl[\sum_{d_i\ge 1}F(d_i-1)\,\alpha^{d_i-1}\Bigr] \label{eq:28}
\end{equation}
We compute the sums in terms of $C_n^q$ (recall that $\card\g\ge
2$ if $\g\in\MM_\mathbb{G}$):
\begin{eqnarray}
\sum_{d_i\ge 1}F(d_i-1)\,\alpha^{d_i-1} &=& \sum_{s_i\ge 2}
C_{s_i}^q \sum_{0\le d_i-1\le s_i} {s_i\choose d_i-1}\a^{d_i-1}
\nonumber\\
&=& \sum_{s_i\ge 2} C^q_{s_i}\,(1+\a)^{s_i}\;. \label{eq:29}
\end{eqnarray}
Likewise
\begin{eqnarray}
\sum_{d_1\ge 1}d_1F(d_1)\,\alpha^{d_1} &=& \sum_{s_1\ge 2}
C_{s_1}^q \sum_{0\le d_1\le s_1} {s_1\choose d_1}\a^{d_1}
\nonumber\\
&=& \sum_{s_i\ge 2} C^q_{s_1}\,\a\,s_1(1+\a)^{s_1-1}\;.
\label{eq:30}
\end{eqnarray}
Finally,
\begin{equation}
B_n\;\le\; {\card{ \mathbb{V}}\a\over n(n-1)}\Bigg[\sum_{s\ge
2}s(1+\a)^{s-1}C^q_{s}\Bigg] \Bigg[{1\over \a}\sum_{s\ge
2}(1+\a)^{s}C^q_{s}\Bigg]^{n-1}\;. \label{eq:31}
\end{equation}
Thus, if
\begin{equation}
{1\over \a}\sum_{s\ge 2}(1+\a)^{s}C^q_{s}\le 1 \label{alf}
\end{equation}
we have, from \reff{eq:ex} and \reff{eq:31},
\begin{equation}
\Sigma^*_{\mathbb{G}}(q)\;=\; \sum_{n\ge 1} B_n \;\le\; \card
{\mathbb{V}}\sum_{s\ge 2}C_s^q \Big[1 +\a s(1+\a)^{s-1}\sum_{n\ge
2}{1\over n(n-1)}\Big] \label{eq:32}
\end{equation}
which is finite, if $q>e\D$, because of the bound \reff{acti}.
Condition \reff{alf} is, in fact, identical to \reff{FP} under the
relabeling $1+\a=e^{a}$.

\subsection{Proof of lemma \ref{lem:3b}}

We combine \reff{2.14} with the bound \reff{eq:10.1} to obtain
\begin{equation}
\sup_{v_0\in \mathbb{V}}\sum_{\g\in \MM_\mathbb{G}:\,\, v_0\in
\g\atop \card\g=n} \bigl|z_\g(q)\bigr| \;\le\;
\frac{1}{q^{n-1}}\sup_{v_0\in \mathbb{V}}\sum_{\g\in
\MM_\mathbb{G} :\,\, v_0\in \g \atop \,\mathbb{G}_\g\;{\rm
connected}} \Bigl| \overline\PP_{\mathbb{G}_\g}\Bigr| \;\le\;
\,\frac{1}{q^{n-1}}\,\bar t_{n}(\D)\;. \label{eq:11}
\end{equation}

\subsection{Proof of Lemma \protect\ref{lem:3}}

Let $U_{v_0}(\D)$ be the infinite tree in which all vertices have
degree $\D$ except for the vertex $v_0$, identified as the root,
which has degree $\D - 1$ (so that  each vertex $v\in U_\D$,
including the root, has $\D-1$ descendants).
Let $\bar u_{n}(\D)$ be   the number  subtrees in $U_{v_0}(\D)$
which have $n$ vertices, contain the root $v_0$, and such that  for
any vertex $v$ of $U_{v_0}(\D)$ and any $k\le \D-1$,
the number $\tilde t_k^G$
defined in \reff{wpen} is
the maximum number of subsets of $k$
descendants of $v$ with fixed cardinality $k$. Define the formal power series

\begin{equation}\label{sU}
\overline U(x)=\sum_{n=1}^\infty   \bar u_{n}(\D)x^n
\end{equation}
From the recursive structure of $(\D-1)$-regular rooted trees we
deduce  that $\overline U(x)$ obey the equations
\begin{equation}\label{U}
\overline U\;=\;x\,\widetilde Z_{\mathbb{G}}(\overline U)
\;=\;\psi_x(\overline U)
\end{equation}
and recalling the definitions (\ref{ZD}),  (\ref{ZD-1}), we also have that the formal
series $\overline T(x)$ defined in (\ref{sTbarb}) is related to $\overline U$ by
\begin{equation}\label{Tbar}
\overline T\;=\;x\,Z_{\mathbb{G}}(\overline U)
\end{equation}
The function
\begin{equation}\label{fU}
f(u)={u \over \widetilde Z_{\mathbb{G}}(u)}
\end{equation}
is zero for $u=0$, increases until a point  $u_0\in [0,\infty]$
where it attains its maximum and then decreases monotonically to
zero in the interval $(u_0,\infty)$. So $f$ is a bijection from
$[0,u_0]$ to $[0,R=f(u_0)]$, with
\begin{equation}
R\;=\;\sup_{u\ge 0}{u\over \widetilde Z_{\mathbb{G}}(u)}
\end{equation}

The function $\psi_x(\overline U)$ defined in \reff{U}, on the other hand,
can be visualized as a sum over single-generation trees where
the root, labelled by $x$, is followed by up to ${\D-1}$ descendants
labelled by $\overline U$.  Hence, its $M$-th iteration, $\psi^M_x(\overline U)$,
corresponds to a sum over a set of $M$-generation trees where all
vertices are labelled by $x$ except those of the $M$-th generation,
which are labelled by $\overline U$.  Applying this argument to $\overline U=u\in
[0,u_0]$ and $x=f(u)\in[0,R]$, we have that
\begin{equation}\label{orly.3}
\sum_{n=1}^M   \bar u_{n}(\D)\,x^n\;\le\; \psi^{M+1}_x(u)\;=\;u\;.
\end{equation}

We conclude that the positive series $\overline
U(x)=\sum_{n=1}^\infty   \bar u_{n}(\D)x^n$ converges for all
$x\in (0,R)$ and furthermore
$$
\overline U^{-1}(u)=f(u) \,\,\,\quad \mbox{for } u\in[0,u_0]\;.
$$
It follows that the
positive series $\overline T(x)=\sum_{n=1}^\infty  \bar t_{n}(\D)
x^n$ converges in the same interval, since $xZ_G(u)\le
x(1+u)\widetilde Z_{G}(u)=u+u^2$ for all $u\in [0,u_0)$.

Finally, we argue:
\begin{eqnarray}
\sup\Bigl\{ 0<x< R:\; \frac{1}{x}\,\overline T(x)\le b \Bigr\}
&=& \sup\Bigl\{ 0<x< R:\; Z_G(\overline U(x))\le b \Bigr\} \\
&=&\sup\Bigl\{ 0<x< R:\; x\le f\big(Z_G^{-1}( b)\big) \Bigr\}\\
&=& \sup\Biggl\{ 0<x< R:\; x\le {Z_G^{-1}( b)\over \widetilde Z_{G}\big(Z_G^{-1}( b)\big)} \Biggr\}\\
&=&{Z_G^{-1}( b)\over \widetilde Z_{G}\big(Z_G^{-1}( b)\big)}\;.\;\qed
\end{eqnarray}

\section*{Acknowledgements}
We thank Alan Sokal for comments and clarification that helped us
to improve our results and its presentation. The work of AP was
supported by a visitor grant of CAPES (Coordena\c{c}\~ao de
Aperfei\c{c}oamento de Pessoal de N\'{\i}vel Superior, Brasil). He
also thanks the Mathematics Laboratory Raphael Salem of the
University of Rouen for the invitation that started the project
and for hospitality during its realization.


\begin{thebibliography}{99}


\bibitem{bor05} C. Borgs (2005): Absence of Zeros for the Chromatic Polynomial
on Bounded Degree Graphs.  Preprint, to appear in {\sl
Combinatorics, Probability and Computing}.

\bibitem{bry84} D. C. Brydges (1984): A short cluster in cluster
  expansions.  In {\sl Critical Phenomena, Random Systems, Gauge
    Theories}, Osterwalder, K.  and Stora, R. (eds.), Elsevier,
  129--83.

\bibitem{cam82} C. Cammarota (1982): Decay of correlations for
  infinite range interactions in unbounded spin systems. {\sl Comm.
    Math. Phys.} {\bf 85}, 517--28.

\bibitem{dob96} R. L. Dobrushin (1996): Estimates of semiinvariants
  for the {I}sing model at low temperatures.  {\sl Topics in
    Statistics and Theoretical Physics}, Amer. Math. Soc. Transl. (2),
  {\bf 177}, 59--81.

\bibitem{grukun71}
C. Gruber and H. Kunz (1971): General properties of polymer
systems. {\sl Comm. Math. Phys.} {\bf 22}, 133--61.

\bibitem{FP} R. Fernandez, A. Procacci (2006): Cluster expansions for
abstract polymer models.  New bounds from an old approach.
Preprint arXiv math-ph/0605041, to appear in {\sl Comm. Math.
Phys}.

\bibitem{kotpre86} R. Koteck\'y and D. Preiss (1986): Cluster
  expansion for abstract polymer models.  {\sl Commun. Math. Phys.},
  {\bf 103}, 491--498.



 \bibitem{pen67} O. Penrose (1967): Convergence of fugacity
 expansions for classical systems.  In {\it Statistical
 mechanics: foundations and applications}\/, A. Bak (ed.),
 Benjamin, New York.

\bibitem{prosco99} A. Procacci, B. Scoppola (1999): Polymer gas approach to
  $N$-body lattice systems.  {\sl J. Statist. Phys.} {\bf 96}, 49--68.

\bibitem{proscoger03} A. Procacci, B. Scoppola and V.Gerasimov (20030:
Potts model on infinite graphs and the limit of chromatic
polynomials. {\sl Commun. Math. Phys.} {\bf 235}, 215--31.

\bibitem{rue69}  D. Ruelle (1969): {\it Statistical mechanics: Rigorous
    results}\/. W. A. Benjamin, Inc., New York-Amsterdam.

\bibitem{sok01} A. Sokal (2001): Bounds on the complex zeros of
(di)chromatic polynomials and Potts-model partition functions.
{\sl Combin. Probab. Comput.} {\bf 10}, 41-77.



\end{thebibliography}
\end{document}